\documentclass[twocolumn,showpacs,preprintnumbers,amsmath,amssymb,superscriptaddress]{revtex4}

\usepackage{graphicx}
\usepackage{dcolumn}
\usepackage{bm}
\usepackage{verbatim}

\begin{document}
\title{Atomic $\check {\textrm{S}}$olc filter: multi-resonant photoemission via periodic poling of atom-cavity coupling constant}

\author{Hyun-Gue Hong}
\author{Wontaek Seo}
\author{Moonjoo Lee}
\author{Younghoon Song}
\affiliation{Department of Physics and Astronomy, Seoul National University, Seoul 151-747, Korea}
\author{Young-Tak Chough}
\address{Department of Optical Communications and Electronic Engineering, Gwangju University, Gwangju 503-703, Korea}
\author{Jai-Hyung Lee}
\author{Kyungwon An}
\email{kwan@phya.snu.ac.kr} 
\affiliation{Department of Physics and Astronomy, Seoul National University, Seoul 151-747, Korea}

\date{\today}

\begin{abstract}
This paper describes a novel atom-cavity interaction induced by periodically poled atom-cavity coupling constant which leads to multiple narrow photoemission bands for an initially inverted two-level atom under the strong coupling condition. The emission bandpass narrowing has a close analogy with the folded $\check{\textrm{S}}$olc filter in the context of quasi-phase matching by periodic poling. We present a closed form solution of the emission probability at the end of interaction and deduce the multiple phase matching condition for this system which is programmable by the
interaction time. The Bloch sphere analysis provides a clear understanding of the underlying atomic dynamics associated with the multiple resonances in the semiclassical limit. Furthermore, we show that this interaction can be applied to generation of the nonclassical field with sub-Poisson photon statistics.\end{abstract}

\maketitle

\section{introduction}

Periodic poling of nonlinear medium \cite{Armstrong1962, Franken1963, Fejer1992, Houe1995} is a widely used technique to achieve quasi-phase matching in nonlinear optics. The phase mismatch otherwise present in the medium is actively compensated by periodically reversing the phase slip so that the nonlinear response of the medium can be maintained over a large distance.
The quasi-phase matching by periodic poling also finds its use in a polarization interference filter, namely the folded $\check {\textrm{S}}$olc filter\cite{Solc1965, Yeh1979}. It is constructed by periodically folding the fast axes of waveplates which are stacked between two orthogonal polarizers. A vertically polarized input is gradually brought into a horizontally polarized output for the quasi-phase matched angle of the fast axes. Since the intensity of the output is strongly dependent on the phase retardation, and in turn on the wavelength of the input field, the transmittance of the system is sharply peaked at target wavelengths. The passband of the filter gets very narrow as the number of stacks increases.

In this paper we describe an atomic $\check {\textrm{S}}$olc filter, an atom-cavity system where a narrowing of photoemission band occurs via a novel atom-cavity interaction under the strong coupling condition. By periodically poling the atom-cavity coupling constant, one can actively guide the initially inverted atom into its ground state, thereby deliver a photon into the cavity mode perfectly at certain atom-cavity detunings satisfying a quasi-phase matching condition. As the number of poling increases the bandwidth for the photoemission gets narrower. We present a closed form solution of the photoemission probability for an arbitrary number of poling. The quasi-phase matching condition, which is programmable by the atom-cavity interaction time, is deduced from the geometric  consideration of the Bloch sphere. We explain the filtering mechanism by examining the trajectories of the atomic state on the Bloch sphere.

The atom traveling across a high-order transverse mode (TEM$_{N0}$) or along the cavity axis in a high-Q Fabry-P\'erot  cavity may well be suited for the realization of the present idea. In particular,  we show that the narrowing of the photoemission probability in the parameter space has a direct application to the generation of micromaser/microlaser field \cite{Meschede1985, An1994} with enhanced sub-Poisson photon statistics \cite{Rempe1990, Choi2006}.

This paper is organized as follows. In Sec.\ \ref{sec_model}, a model Hamiltonian and a periodic atom-cavity coupling constant are introduced. Time domain analysis in Sec.\ \ref{sec_pem} provides the emission probability in a closed form by using the unitary evolution matrix. In Sec.\ \ref{sec_bloch}, the regular structure observed in the emission probability and the phase matching conditions therein are interpreted in terms of the Bloch sphere. The analogy between the conventional $\check {\textrm{S}}$olc filter and our system is given in Sec.\ \ref{sec_solc}. Possible applications to nonclassical light generation and a frequency domain analysis are presented in Sec.\ref{sec_subpoisson}. We conclude in Sec.\ \ref{sec_conclusion}.

\section{model}\label{sec_model}

\begin{figure}
\centering
\includegraphics[width=3.4in]{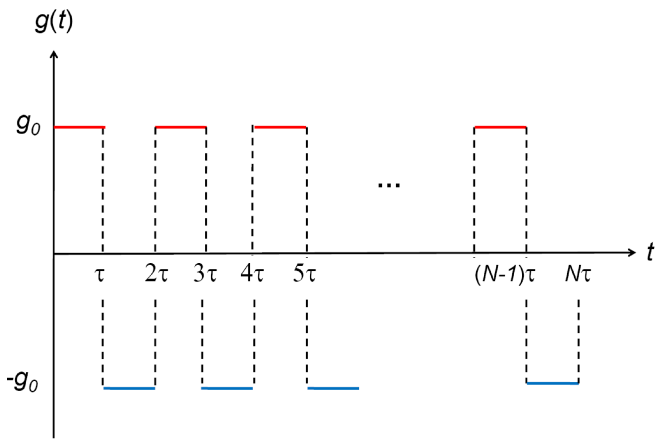}
\caption{Periodically poled atom-cavity coupling constant shown as a function of time.} \label{gt}
\end{figure}

We consider a two-level atom interacting with an electromagnetic field mode of a cavity via electric dipole interaction. Under the rotating wave approximation and in the semiclassical regime of a large mean photon number $n (\gg 1)$ in the cavity mode, the interaction Hamiltonian is given by
\begin{equation}
H^{(m)}_I=\hbar \sqrt{n} g (\sigma_+e^{-i\Delta t}+\sigma_-e^{i\Delta t})\;,
\end{equation}
where $g$ is the atom-cavity coupling constant, $\sigma_{\pm}$ are the Pauli pseudo-spin operators and $\Delta$ is the cavity-atom detuning. In our model system, the relative phase of the atom-cavity coupling constant is sequentially changing by $\pi$ after each time interval
of $\tau$, {\em i.e.},
\begin{equation}
g(t)=(-1)^{m+1}g_0 \qquad\textrm{for}\quad (m-1)\tau<t<m\tau \;,
\label{couplingEq}
\end{equation}
where  $m$=1, 2, 3, \ldots, $N$ (see Fig.\ \ref{gt}). We call such an interaction the $N$th-order multipolar interaction. The equation of motion is then
\begin{equation}
{\dot c_e}(t)=-i\sqrt{n} g(t)e^{-i\Delta t} c_g(t)\;,
\label{equationMotion1}
\end{equation}
\begin{equation}
{\dot c_g}(t)=-i\sqrt{n}g(t)e^{i\Delta t} c_e(t)\;,\;\;
\label{equationMotion2}
\end{equation}
where $c_e(t)$ and $c_g(t)$ are the probability amplitude of the excited and ground state, respectively, in the interaction picture. Our interest is to find the photoemission probability of an initially inverted atom after the entire interaction time of $N\tau$.

Although the equations are written in the form of semiclassical Bloch equation, this model is also applicable to the vacuum Rabi oscillation in the restricted atom-field dressed basis of $|e,0\rangle$ and $|g,1\rangle$ with $\sqrt{n}$ removed and $2g_0$ understood as the vacuum Rabi frequency. Then the time evolution of the atomic system is exactly of the same form for both quantum and semiclassical fields if we trace out the photonic variables in the quantum calculation.

\section{Emission probability}\label{sec_pem}
If we denote the atomic wave function in the interaction picture as a column vector, the evolution of the atomic state is described by a unitary matrix as
\begin{eqnarray}
\left[ \begin{array}{cc}
c^{(m)}_{e} \\
c^{(m)}_{g}
\end{array} \right]=U^{(m)}\left[ \begin{array}{cc}
c_{e}^{(m-1)} \\
c_{g}^{(m-1)}
\end{array} \right]\;,\\\nonumber
\end{eqnarray}
where $c^{(m)}_e$ and $c^{(m)}_g$ are the excited and ground state probability amplitude at $t=m\tau$, respectively. 
The elements of $U^{(m)}$ are given by
\begin{eqnarray}
\left(U^{(m)}\right)_{11}&=&\left(\cos{\frac{\phi}{2}}-i\frac{\delta}{\phi}\sin{\frac{\phi}{2}}\right)e^{i\delta/2}=\left(U^{(m)}\right)^*_{22}\nonumber\\
\left(U^{(m)}\right)_{12}&=&(-1)^m i \frac{\eta}{\phi}\sin{\frac{\phi}{2}} e^{i(m-1/2)\delta}\nonumber\\
&=&-\left(U^{(m)}\right)^*_{21}\label{umat}
\end{eqnarray}
where $\eta=2\sqrt{n}g_0\tau$, $\delta=\Delta\tau$, and $\phi=\sqrt{\eta^2+\delta^2}$, the Rabi precession angle. As mentioned in Sec.\ \ref{sec_model}, this matrix is same for the vacuum Rabi oscillation and the semiclassical one. Note also that we should keep the index $m$ indicating the sequential order in order to incorporate the correct initial conditions for each stage since $[H_I(t_1), H_I(t_2)]\neq 0$ in general.

We are now interested in the probability by which an initially inverted atom emits a photon into the cavity mode. The probability, called photoemission probability $P_{em}$, is nothing but the ground state probability $|c^{(N)}_g|^2$ of the atom after the entire interaction time $N\tau$. For calculating $P_{em}$ it is convenient to define a unit cell operation which is comprised of two intervals with $\pm g_0$ as
\begin{equation}
T^{(k)}=U^{(2k)}U^{(2k-1)}
\end{equation}
where its elements are
\begin{eqnarray}
\left(T^{(k)}\right)_{11}&=&\left(1-\frac{2\delta^2}{\phi^2}\sin^2{\frac{\phi}{2}}-i\frac{\delta}{\phi}\sin\phi\right)e^{i\delta}=\left(T^{(k)}\right)^*_{22}\nonumber\\
\left(T^{(k)}\right)_{12}&=&-2\frac{\eta\delta}{\phi^2}\sin^2{\frac{\phi}{2}} e^{-i\delta} e^{2ik\delta}=-\left(T^{(k)}\right)^*_{21}.
\end{eqnarray}
For even $N$ the final state is then simply
\begin{eqnarray}
\left[ \begin{array}{cc}
c^{(N)}_{e} \\
c^{(N)}_{g}
\end{array} \right]=\prod_{k=1}^{N/2}T^{(k)}\left[ \begin{array}{cc}
c_{e}^{(0)} \\
c_{g}^{(0)}
\end{array} \right]
\end{eqnarray}
and for odd $N$
\begin{eqnarray}
\left[ \begin{array}{cc}
c^{(N)}_{e} \\
c^{(N)}_{g}
\end{array} \right]=U^{(N)}\prod_{k=1}^{(N-1)/2}T^{(k)}\left[ \begin{array}{cc}
c_{e}^{(0)} \\
c_{g}^{(0)}
\end{array} \right].
\end{eqnarray}

\begin{widetext}
\begin{center}
\begin{table}
\caption{Emission probabilities after the entire interaction time $N\tau$}
\centering
\begin{tabular}{|c|c|c|c|}
\hline\hline $N$ & $P_{em}^{(N)}/P_{em}^{(1)}$ where $P_{em}^{(1)}=\frac{\eta^2}{\phi^2}\sin^2{\frac{\phi}{2}}$ & $N$ & $P_{em}^{(N)}/P_{em}^{(2)}$ where $P_{em}^{(2)}=\frac{4\eta^2\delta^2}{\phi^4}\sin^4{\frac{\phi}{2}}$\\
[0.5ex] \hline
3 & $\left[ 1-4\frac{\delta^2}{\phi^2}\sin^2{\frac{\phi}{2}}\right]^2$ & 4 & $\left[ 2-4\frac{\delta^2}{\phi^2}\sin^2{\frac{\phi}{2}}\right]^2$ \\
5 & $\left[ 1-12\frac{\delta^2}{\phi^2}\sin^2{\frac{\phi}{2}}+16\frac{\delta^4}{\phi^4}\sin^4{\frac{\phi}{2}}\right]^2$ & 6 &$\left[ 3-16\frac{\delta^2}{\phi^2}\sin^2{\frac{\phi}{2}}+16\frac{\delta^4}{\phi^4}\sin^4{\frac{\phi}{2}}\right]^2$ \\
7 & $\left[
1-24\frac{\delta^2}{\phi^2}\sin^2{\frac{\phi}{2}}+80\frac{\delta^4}{\phi^4}\sin^4{\frac{\phi}{2}}-64\frac{\delta^6}{\phi^6}\sin^6{\frac{\phi}{2}}\right]^2$
& 8 & $\left[
4-40\frac{\delta^2}{\phi^2}\sin^2{\frac{\phi}{2}}+96\frac{\delta^4}{\phi^4}\sin^4{\frac{\phi}{2}}-64\frac{\delta^6}{\phi^6}\sin^6{\frac{\phi}{2}}\right]^2$
\\ [1ex] \hline
\end{tabular}
\label{table:pem}
\end{table}
\end{center}
\end{widetext}

\begin{figure}
\centering
\includegraphics[width=3.4in]{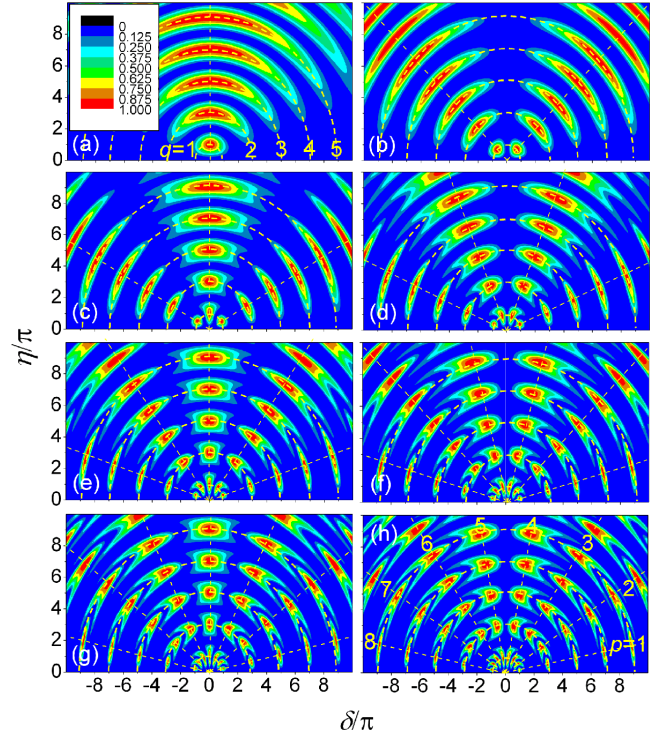}
\caption{Density plot of emission probability in $\delta$-$\eta$ plane. From (a) to (h) $N$ increases from 1 to 8. The dashed lines (called branch lines) indicate the maximum visibility of Rabi oscillation and designated by the index $p$ in the text. We can also recognize the semicircular patterns (called branch circles), which are designated by $q$. Quasi-phase matching is accomplished at the intersections of the branch lines and the branch circles.}
\label{map}
\end{figure}

Note that each matrix element of $T^{(k)}$ is independent of the sequential order $k$ except the extra phase factors $e^{\pm2ik\delta}$. We can show that those phase factors do not affect the \textit{absolute} value of the probability amplitude (see Appendix A) so that we may eliminate them from $T^{(k)}$ and define the resulting matrix ${\cal T}$ as
\begin{eqnarray}
{\cal T}_{11}&=&1-\frac{2\delta^2}{\phi^2}\sin^2{\frac{\phi}{2}}-i\frac{\delta}{\phi}\sin\phi={\cal T}^*_{22}\nonumber\\
{\cal T}_{12}&=&-2\frac{\eta\delta}{\phi^2}\sin^2{\frac{\phi}{2}}=-{\cal T}^*_{21}.\label{Tmat}
\end{eqnarray}
This elimination of $k$ dependence reduces the complexity of the calculation much. From the Chebyshev identity \cite{Yariv1977} the product of an unitary unimodular matrix is given as
\begin{eqnarray}
&&\left[ \begin{array}{cc}
A & B\\
C & D
\end{array} \right]^{N}\nonumber\\&&=\left[ \begin{array}{cc}
\frac{A\sin N\xi-\sin(N-1)\xi}{\sin\xi} & \frac{B\sin N\xi}{\sin\xi}\\
\frac{C\sin N\xi}{\sin\xi} & \frac{D\sin N\xi-\sin(N-1)\xi}{\sin\xi}
\end{array} \right]
\end{eqnarray}
where $\cos\xi=\frac{1}{2}(A+D)$.
Hence, the final results for an initial state
\begin{eqnarray}
\left[ \begin{array}{cc}
c^{(0)}_{e} \\
c^{(0)}_{g}
\end{array} \right]=\left[ \begin{array}{cc}
1 \\
0
\end{array} \right]
\end{eqnarray}
are for even $N=2k$ ($k$=1, 2, 3, \ldots)
\begin{eqnarray}
|c_g^{(N)}|^2&=&\left[\left({\cal
T}^{k}\right)_{21}\right]^2\nonumber\\&=&\frac{4\eta^2\delta^2}{\phi^4}\left(\frac{\sin
k\xi}{\sin\xi}\right)^2\sin^4{\frac{\phi}{2}}
\end{eqnarray}
and for odd $N=2k+1$ ($k$=1, 2, 3, \ldots)
\begin{eqnarray}
|c_g^{(N)}|^2&=&\left|\left(U^{(N)}\prod_{j=1}^{k}T^{(j)}\right)_{21}\right|^2 \nonumber\\
&=&\frac{\eta^2}{\phi^2}\sin^2{\frac{\phi}{2}}\biggl[\left(1-4\frac{\delta^2}{\phi}\sin^2{\frac{\phi}{2}}
\right)\frac{\sin k\xi}{\sin\xi}\nonumber\\&&-\frac{\sin
(k-1)\xi}{\sin\xi} \biggr]^2\;,
\end{eqnarray}
where
\begin{equation}
\cos\xi=1-\frac{2\delta^2}{\phi^2}\sin^2{\frac{\phi}{2}}.
\end{equation}
Using the Chebyshev polynomial in the trigonometric form
\cite{ArfkenBook} one can expand
\begin{eqnarray}
\frac{\sin k\xi}{\sin\xi}&=&\biggl[ \binom{k}{1}\cos^{k-1}{\xi}-\binom{k}{3}\cos^{k-3}\xi\sin^2\xi\nonumber\\&&+\binom{k}{5}\cos^{k-5}\xi\sin^{4}\xi+\cdots\biggr]
\end{eqnarray}
to have the emission probability as a function of Rabi frequency ($\eta$) and atom-cavity detuning ($\delta$). First few of them are explicitly listed in Tab.\ref{table:pem} and plotted in Fig.\ \ref{map}.

The result for $N=1$ is the familiar Rabi oscillation (Fig.\ \ref{map}(a)). The visibility of the oscillation is maximum on resonance ($\delta=0$). For $N>1$ the additional regular patterns emerge in the $(\eta, \delta)$ map as in Fig.\ \ref{map}(b)-(h). In particular, the 100\% certain emission of a photon is possible for off-resonant conditions. We notice full visibility along a specific straight line, {\em i.e.} at a fixed ratio of $\eta$ to $\delta$. Furthermore, the fringe pattern gets sharper for the smaller $\eta/\delta$ ratio. In the next sections we investigate what physical processes are involved for such patterns and the sharpness of the fringes.

\section{Bloch sphere analysis}\label{sec_bloch}

\begin{figure}
\centering
\includegraphics[width=3.4in]{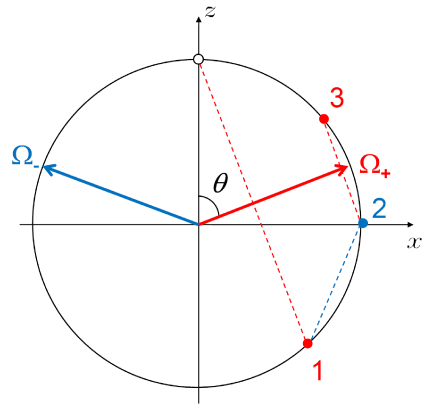}
\caption{Evolution of the Bloch vector in $x-z$ plane assuming $\phi=\pi$. The filled circles are the location of the Bloch vector after each $\tau$ and ${\bf \Omega_{\pm}}$ are torque vectors associated with $\pm g_0$, respectively.}
\label{findingtheta}
\end{figure}

The coherent evolution of the internal state of a two-level atom is easily visualized as the rotational motion of a state vector around a torque vector on the Bloch sphere in the semiclassical limit \cite{Feynman1957}. The equation of motion is
\begin{equation}
{\dot {\bf R}}={\bf R}\times{\bf \Omega}
\end{equation}
where ${\bf R}=\left[\frac{1}{2}\left(c_ec^*_g+c_gc^*_e\right),
\frac{i}{2}\left(c_ec^*_g-c_gc^*_e\right), |c_e|^2-|c_g|^2\right]$ the state vector, and ${\bf \Omega}=\left[2\sqrt{n}g(t), 0, \Delta\right]$ the torque vector. Initially the state vector $\bf{R}$ is pointing the north pole. When the interaction is turned on, ${\bf R}$ precesses around the torque vector $\bf{\Omega}$. Flipping the polarity of the coupling constant is equivalent to fold the torque vector with respect to $z$-axis in the $x$-$z$ plane (as in ${\bf \Omega_+}$ and ${\bf \Omega_-}$ in Fig.\ \ref{findingtheta}).

\begin{figure}
\centering
\includegraphics[width=3.4in]{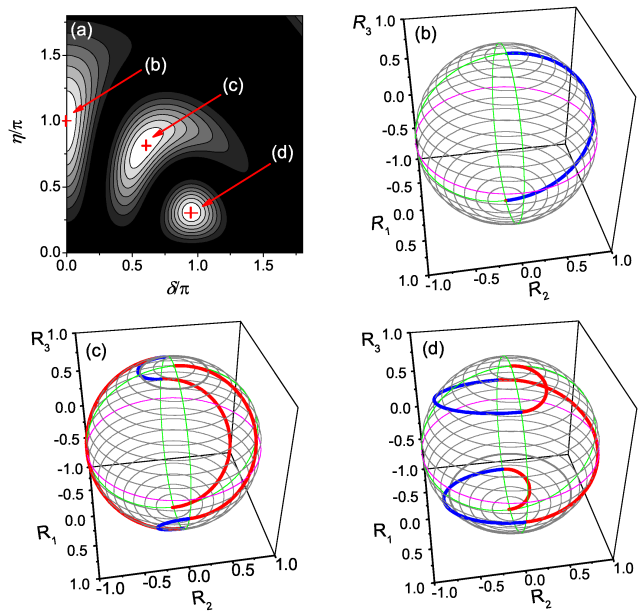}
\caption{(a) Emission probability near $\phi=\pi$ ($q=1$) semicircular branch for $N=5$. The red crosses indicate the points for the complete emission of a photon. The corresponding trajectories of the Bloch vector are shown for (b) $p=3$, (c) $p=2$ and (d) $p=1$. The red curves represent the evolution under $+g_0$ while the blue curves are under $-g_0$ as in the same color convention of Fig.\ \ref{gt}.}
\label{trajectory}
\end{figure}

Suppose the cavity is tuned on resonance, corresponding to the line of $\delta=0$ in Fig.\ \ref{map}. Since ${\bf \Omega}$ is pointing the exactly opposite directions periodically, the state vector just retraces a part of the great circle as in Fig.\ \ref{trajectory}(b). Thus for even $N$ the interaction ends up with no emission of a photon. We see the emission probabilities are completely quenched along $\delta=0$ in Fig.\ \ref{map} for even $N$. For odd $N$, on the other hand, the interaction is effectively unipolar since after $(N-1)\tau$ the state vector comes back to its initial position. Indeed along $\delta=0$ the emission probabilities are the same for all odd $N$ as shown in Fig.\ \ref{map}.

For off-resonance, however, the multistep flipping of the torque vector brings about various trajectories on the Bloch sphere. Some of such nontrivial motions of the state vector lead to complete emission of a photon even at off-resonance unlike the conventional unipolar interaction. In the $\delta$-$\eta$ plane of Fig.\ \ref{map} these occur only along some specific ratios of $\eta/\delta$, {\em i.e.} at specific angles of the torque vector. Some of such trajectories of the state vector are shown in Figs.\ \ref{trajectory}(c) and \ref{trajectory}(d). The trajectory with the smallest $\eta/\delta$ [Fig.\ \ref{trajectory}(d)] is noticeable among them in that it steadily goes down to the ground state, utilizing all $N$ steps of interaction in increasing the emission probability. In the other trajectories, on the other hand, the interactions in some intervals are contributed to excite the atom back although the state vector eventually arrives at the ground state in the end.

The requirement for the complete emission, or namely the quasi-phase matching condition, can be found via a simple geometric argument. Assume the accumulated net precession angle $\phi$ in Eq.\ (\ref{umat}) is odd multiples of $\pi$ for each interval. This condition constitutes the semicircles, called branch circles, in the map of $P_{em}$ in Fig.\ \ref{map}. Then the atomic state after each interval reduces onto the $x$-$z$ plane of Bloch sphere as in Fig.\ \ref{findingtheta}, where the position of the state vector is numbered for a few sequence of interaction.

In ($x$, $z$) coordinate a reflection about the torque vector is represented by
\begin{eqnarray}
& &\left[ \begin{array}{cc}
\cos\theta & (-1)^{m+1}\sin\theta\\
(-1)^{m}\sin\theta & \cos\theta
\end{array} \right]
\left[ \begin{array}{cc}
-1 & 0\\
0 & 1
\end{array} \right]\nonumber\\&&
\left[ \begin{array}{cc}
\cos\theta & (-1)^{m}\sin\theta\\
(-1)^{m+1}\sin\theta & \cos\theta
\end{array} \right] \nonumber\\
& &=\left[ \begin{array}{cc}
-\cos2\theta & (-1)^{m+1}\sin2\theta\\
(-1)^{m+1}\sin2\theta & \cos2\theta
\end{array} \right]
\end{eqnarray}
for $(m-1)\tau<t<m\tau$ with $\theta$ being the angle between $z$-axis and the torque vector. Then the unit cell operation of $N=2$ is represented by
\begin{eqnarray}
\left[ \begin{array}{cc}
\cos4\theta & -\sin4\theta\\
\sin4\theta & \cos4\theta
\end{array} \right]
\end{eqnarray}
and the generalization to an arbitrary even $N$ is straightforwardly given as
\begin{eqnarray}
\left[ \begin{array}{cc}
\cos2N\theta & -\sin2N\theta\\
\sin2N\theta & \cos2N\theta
\end{array} \right].\label{refMatEven}
\end{eqnarray}
Applying one more reflection to Eq.(\ref{refMatEven}) gives the odd $N$ result which turns out to be
\begin{eqnarray}
\left[ \begin{array}{cc}
-\cos2N\theta & \sin2N\theta\\
\sin2N\theta & \cos2N\theta
\end{array} \right].
\end{eqnarray}
Therefore the position of the state in ($x,z$) coordinate changes as
\begin{eqnarray}
\left[ \begin{array}{cc}
0\\
1
\end{array} \right]\to
\left[ \begin{array}{cc}
(-1)^{N+1}\sin2N\theta\\
\cos2N\theta
\end{array} \right].
\end{eqnarray}
Thus the quasi-phase matching occurs for $\cos2N\theta=-1$, that is,
\begin{equation}
\theta=\left(\frac{p-1/2}{N}\right)\pi\;,
\end{equation}
where $p=1, 2, \cdots, N$. These particular angles constitute the straight lines, called branch lines, in Fig.\ \ref{map}. Since $\eta=\delta\tan\theta$ and $\phi=\sqrt{\eta^2+\delta^2}=(2q-1)\pi$, where $q=1,2,\cdots$, the optimal detuning for complete photoemission is given by
\begin{equation}
\delta_{opt}/\pi=\pm(2q-1)\cos\left[\left(\frac{p-\frac{1}{2}}{N}\right)\pi \right]\;,
\label{optdelta}
\end{equation}
where $p$ and $q$ index the branch lines and the branch circles, respectively.

\section{$\check {\textrm{S}}$olc filter analogy}\label{sec_solc}

It is noted in Fig.\ \ref{map} that the widths of fringes along different branch lines are significantly different from each other. The fringes are sharper along a branch line with smaller $|\theta|$. As mentioned in the previous section, the state-vector trajectories associated with different $p$'s are qualitatively different from each other. In particular, precise multiple rotations are sequentially involved for the trajectory of Fig.\ \ref{trajectory}(d), and thus a slight slip from the optimal condition will lead to a significant reduction in the photoemission probability. That is why the fringe along the branch line of $p=1$ (or equivalently $p=N$) is the sharpest. Furthermore, the passband of $\delta$ gets narrower as $N$ gets large.

This behavior is analogous to what happens in the optical $\check{\textrm{S}}$olc filter, where the waveplates with their fast axes inclined by $\pm \theta$ about the input vertical polarization are periodically stacked and the phase retardation angle $\phi$ imposed by each waveplate is set to $\pi$ (half wave plate) for a target wavelength. Here the angles $\theta$ and $\phi$ in the optical $\check{\textrm{S}}$olc filter work as nothing but the inclined angle of the torque vector and the net precession angle, respectively, in our atom-cavity case while the two kinds of polarization correspond to the two energy states of the atom and the Jones matrix of the filter corresponds to the matrix ${\cal T}$.

\begin{figure}
\centering
\includegraphics[width=3.4in]{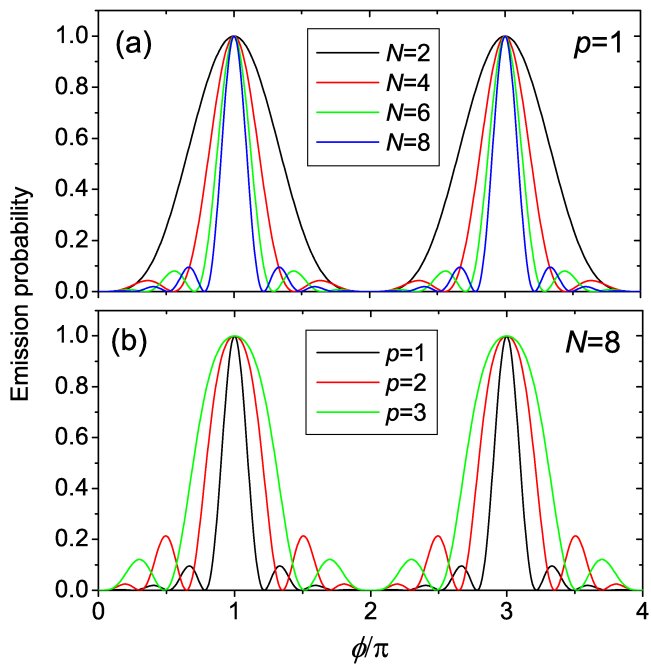}
\caption{(a) For given phase matching order $p=1$ (or $\theta=\pi/2N$), the passband of the atomic $\check {\textrm{S}}$olc filter, our atom-cavity system, is shown as a function of $\phi(\eta, \delta)$. (b) For given number of poling $N=8$, the passband for different phase matching order $p$ are shown.}
\label{filterSpectrum}
\end{figure}

In fact, one-to-one correspondence can be explicitly shown if we restrict ourselves to the absolute value of the probability amplitude $c_{e,g}(N\tau)$. If we perform the coordinate transformation
\begin{eqnarray}
\left\{ \begin{array}{ll}
\theta=\tan^{-1}(\eta/\delta) \\
\phi=\sqrt{\eta^2+\delta^2}
\end{array} \right.
\end{eqnarray}
we can obtain the spectrum of filtering (actually passband of photoemission) of our atom-cavity system as a function of $\phi$ for given azimuth $\theta$ as in the optical $\check{\textrm{S}}$olc filter. We confirm two aspects of filtering functions in Fig.\ \ref{filterSpectrum}. Firstly the passband gets narrower for larger $N$ and secondly for given $N$ the lowest $p$ (or equivalently the largest $p$) gives the sharpest passband of photoemission. Due to this analogy our atom-cavity system can be viewed as an atomic $\check{\textrm{S}}$olc filter.

However, we also recognize the difference between the two systems when we take the phase of the probability amplitude into account such as the atomic polarization, for which the optical $\check{\textrm{S}}$olc filter has no corresponding counterpart. Note then we have to use the phase sensitive matrix $T^{(k)}$ instead of ${\cal T}$ to retrieve the phase information in our atom-cavity system.

\section{discussion}\label{sec_subpoisson}

\subsection{Sub-Poisson photon statistics}

The squeezed lineshape of $P_{em}$ in $\delta-\eta$ plane, brought about by the multi-polar interactions, is useful for generation of nonclassical field state with sub-Poissonian photon statistics in the cavity-QED micromaser/microlaser. While the mean photon number of the conventional laser is stabilized by the cavity damping only \cite{Choi2006}, the active role of coherent medium is also very important in the cavity-QED lasers. Specifically, the negative slope of $P_{em}$ as a function of the mean photon number $n$ reduces the recovery time of the system when the intracavity photon number is momentarily deviated from its stationary point \cite{Choi2006}, and thereby can reduce the photon number fluctuation below the shot-noise limit, resulting in a sub-Poissonian photon statistics. On the contrary, the slope of $P_{em}$ is always positive in the conventional laser. As a measure of the photon number stabilization, we use the Mandel $Q$ factor, which can be written as

\begin{equation}
Q\simeq\frac{\partial P_{em}/\partial n}{D_{\rm cav}-\partial P_{em}/\partial n}
\end{equation}
where $D_{\rm cav}$ is given by the cavity decay rate divided by the atomic flux \cite{Choi2006}. The negative value of $Q$ indicates a sub-Poissonian photon statistics.

\begin{figure}
\centering
\includegraphics[width=3.4in]{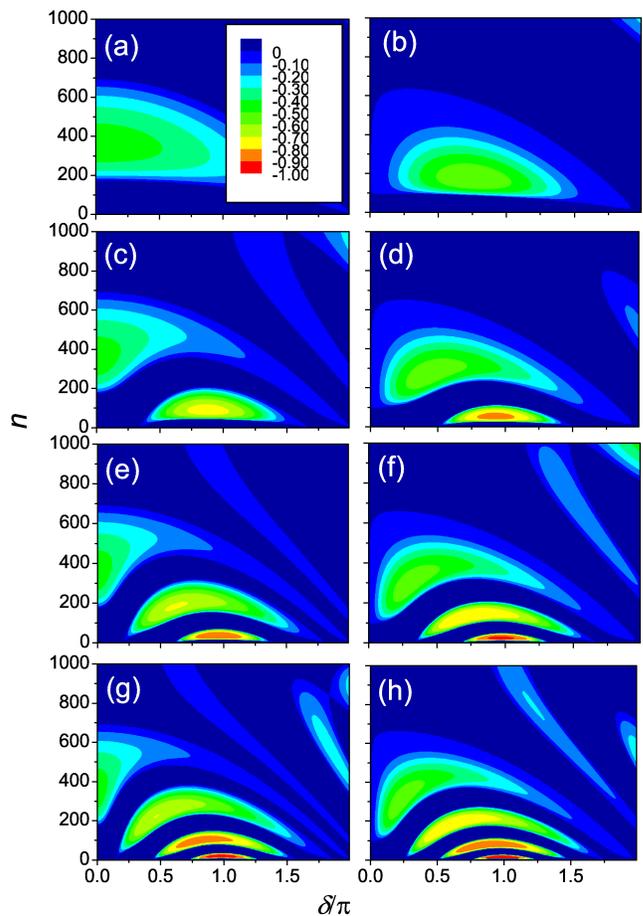}
\caption{Mandel Q factor for $\eta_0$=0.24 and $D_{\rm cav}$=0.0038 as $N$ is varied from 1 (a) to 8 (h). Only the negative part is shown for simplicity.}
\label{mandelQ}
\end{figure}

We plot the Mandel $Q$ in Fig.\ \ref{mandelQ} for various $N$ in $n-\delta$ plane with $\eta_0=2g_0\tau$ fixed at 0.24 and $D_{\rm cav}$=0.0038, both of which are typical experimental parameters of Ref.\ \cite{Choi2006}. Due to the squeezed structure of $P_{em}$ for large $N$'s, we expect the more negative slope of $P_{em}$ for the higher-order multipolar interaction. However, the direction of squeezing is along the branch lines, not along $n$, except for the first branch circle, along which the squeezing is also pronounced as shown in Fig.\ \ref{map}. Therefore, the narrowing effect along $n$ is only good for small detuning of $\delta/\pi \sim \pm1$ for large $N$. For large detunings the direction of squeezing is roughly along $\delta$ for the first and the last branch lines ($p=1, N$). We find the first/last branch line solution with $\phi=\pi$ ($q=1$) provides the most negative slope of $P_{em}$ along $n$ for larger $N$, and thus the most negativity of $Q (\simeq -1)$ or the most nonclassicality of the field as shown in Fig.\ \ref{mandelQ}.

In an experiment employing an atomic beam with a velocity spread, the slope of $P_{em}$ would be reduced due to velocity averaging. However, the peak giving rise to the most negative slope ($p=\{1, N\}, q=1$) is least affected by this averaging and thus its Mandel $Q$ is expected to be somewhat robust to such statistical averaging.

\subsection{Fourier analysis and large $N$ limit}

The coupling constant in Eq.\ (\ref{couplingEq}) is a truncated periodic square-pulse train and can be expressed by a Fourier series on the time interval $[0, N\tau]$ as
\begin{equation}
g(t)=\sum_{l=-\infty}^{\infty}{\cal G}^{(l)}e^{-i2\pi lt/N\tau}
\end{equation}
where
\begin{eqnarray}\setlength\arraycolsep{2pt}
{\cal G}^{(l)}=\frac{1}{N\tau}\int_{-\infty}^{\infty}g(t)e^{i2\pi lt/N\tau}dt.
\end{eqnarray}
Likewise the probability amplitudes can be written in the similar form
\begin{equation}
c_{e,g}(t)=\sum_{l=-\infty}^{\infty}{\cal C}_{e,g}^{(l)}e^{-i2\pi lt/N\tau}.
\end{equation}
The differential equations Eq.\ (\ref{equationMotion1}) and Eq.\ (\ref{equationMotion2}) are transformed into algebraic equations of ${\cal C}^{(l)}_{e,g}$ as
\begin{equation}
{\cal C}^{(l)}_{e,g}=\frac{N\eta}{2\pi l}\sum_{k=-\infty}^{\infty}\frac{{\cal G}^{(l-k\pm s)}}{2g_0} {\cal C}_{g,e}^{(k)}\;,
\label{secular}
\end{equation}
respectively, where we index the detuning with an integer $s$ as $\Delta=2\pi s/N\tau$. Initially only ${\cal C}^{(0)}_e$ is nonzero. As the interaction proceeds the Fourier elements ${\cal G}^{(l)}$ of the coupling constant generates the ground state probability amplitude.

For small $\eta<1$, so called the weak coupling limit, the Fourier coefficients are coupled approximately in the first order of $\eta$. Since the precession angle is very small, ${\cal C}_e^{(l)}$ almost stays in its initial value and ${\cal C}_e^{(0)}$ is dominant over the summation in Eq.\ (\ref{secular}). Then ${\cal C}_g^{(0)}$, which is the dominant frequency components of the ground state amplitude, resonantly grows for the frequencies at which ${\cal G}^{(s)}$ is significant. Therefore, the $\delta$-dependence of $P_{em}$ almost follows the spectral lineshape of $g(t)$ in the first order approximation. Further occupation of ${\cal C}_g^{(l)}$ is accomplished for the strong coupling condition ($\eta>1$) and the rich structure of $P_{em}$ is then revealed.


In the limit of large $N$, the coupling constant is well approximated by the sinusoidally modulated function which in turn leads to the bichromatic atom-field interaction. If the modulation frequency ($\pi/\tau$ here) is much larger than the Rabi frequency, the system is well described by the shifted resonance \cite{Hong2009}. In fact, for $\eta\ll 1$ the emission probability is peaked at $\delta\simeq\pm\pi$, which are shifted resonance frequencies. This claim becomes more correct as $N$ increases as we see at the very bottom of each map in Fig.\ \ref{map} because the coupling constant is more like a sinusoidal function. For $\eta>1$ it is known that the harmonic resonances come in to play so that multiple branches of peaks in the emission probability emerges as in Fig.\ \ref{map}. Although the bichromatic interaction has been extensively studied \cite{Hong2009, Silverans1985, Eberly1988, Mossberg1989}, the systematic analysis for the finite train of flipping Rabi frequency as given here never appeared before.

\section{conclusion}\label{sec_conclusion}

We have studied the atom-cavity system whose coupling constant is periodically poled $N$ times in time. We have shown that the photoemission probability of an initially inverted two-level atom at the end of the interaction exhibits narrowed passbands determined by $\check{\textrm{S}}$olc filter-type quasi-phase matching condition. The closed form expression for the emission probability as a function of Rabi frequency and detuning is provided for an arbitrary number of poling. From the Bloch sphere interpretation we can verify the underlying physical processes of the quasi-phase matching in the emission probability. The narrowing of the photoemission probability in the parameter space can be applied to the generation of micromaser/microlaser field with enhanced sub-Poisson photon statistics.

\begin{acknowledgments}
 This work was supported by NRL and WCU Grants.
\end{acknowledgments}

\appendix
\section{phase factors in the unit cell operation}

The transfer matrix $T^{(k)}$ is a unitary unimodular matrix. Along with the particular form of its phase factors it can be written in its most general form as
\begin{eqnarray}
T^{(k)} =
\left[ \begin{array}{cc}
ae^{i\theta} & be^{i(2k-1)\theta}  \\
-b^*e^{-i(2k-1)\theta} & a^*e^{-i\theta}  \\
\end{array} \right]
\end{eqnarray}
where $a$ and $b$ are arbitrary complex numbers satisfying the unimodular relation $|a|^2+|b|^2=1$. The product of such matrices is of the form
\begin{eqnarray}
\prod_{k=1}^{N}T^{(k)} =
\left[ \begin{array}{cc}
f(a,b)e^{iN\theta} & g(a,b)e^{iN\theta}  \\
-g^*(a,b)e^{-iN\theta} & f^*(a,b)e^{-iN\theta}  \\
\end{array} \right]\label{productMatrix}
\end{eqnarray}
where the functions $f(a,b)$ and $g(a,b)$ are some polynomial functions of $a$ and $b$. The emission probability corresponds to $|g(a,b)|^2$, in which the common phase factors do not appear.

\end{document}